\documentclass[nofootinbib,10pt]{revtex4}
\usepackage{graphicx}
\usepackage{psfrag}

\newcommand{\Ref}[1]{(\ref{#1})}

\def \subs{\subsection}

\newcommand{\mat}[2]{\left(\begin{array}{#1} #2 \end{array}\right)}
\def \f{\frac}

\let \pp=\partial
\newcommand{\beq}{\begin{equation}}
\newcommand{\eeq}{\end{equation}}
\newcommand{\beqs}{\begin{eqnarray}}
\newcommand{\eeqs}{\end{eqnarray}}

\def \wtl{\widetilde}
\def \N{\mathrm{I\hspace{-.37ex}N}}
\def \what{\widehat}

\def \A{{\cal A}}
\def \S{{\cal S}}

\def \ra{\rangle}

\begin{document}
\thispagestyle{empty}

\title{\Large
Spectra of Length and Area \\
in $(2+1)$ Lorentzian Loop Quantum Gravity}

\author{\vspace{3mm}
Laurent Freidel\,${}^{ab}$\footnotemark
\footnotetext{freidel@ens-lyon.fr}
, Etera R. Livine\,${}^{c}$\footnotemark
\footnotetext{livine@cpt.univ-mrs.fr}
, Carlo Rovelli\,${}^{cd}$\footnotemark
\footnotetext{rovelli@cpt.univ-mrs.fr}
}

\affiliation{\vspace{2mm}  ${}^{a}$ Perimeter Institute for Theoretical
Physics, 35 King street North,
Waterloo  N2J-2G9, Ontario, Canada\\
 ${}^{b}$Laboratoire de Physique, {\'E}cole Normale
Sup{\'e}rieure de Lyon, 46 al.~d'Italie, F-69364 Lyon Cedex
07, France \\
 ${}^{c}$ Centre de Physique Th{\'e}orique, Luminy, Case 907, 13288 Marseille Cedex 09, France
\\
 ${}^{d}$ Dipartimento di Fisica, Universit{\'a} di Roma ''La Sapienza", P A Moro 2, I-00185, Roma, Italy}

\date{\today}

\begin{abstract}

\vskip2mm
\centerline{\bf Abstract}

We study the spectrum of the length and area operators in
Lorentzian loop quantum gravity, in 2+1 spacetime dimensions.
We find that the spectrum of {\em spacelike} intervals is {\em continuous},
whereas the spectrum of {\em  timelike} intervals is {\em discrete}.
This result contradicts the expectation that spacelike intervals
are always discrete. On the other hand, it is consistent with the
results of the spin foam quantization of the same theory.

\end{abstract}

\maketitle

\vskip1cm

 \section{Introduction}

A characteristic feature of the loop approach to quantum gravity
is the discrete spectrum of several geometrical quantities.  In
four spacetime dimensions (4d), the easiest geometrical operator
to diagonalize is the area, and its eigenvalues turn out to be
discrete \cite{discrete1,discrete2}.  This is generally expected
to be true in the Euclidean as well as in the Lorentzian case,
since the two theories can be formulated using the same
kinematics, differing only in the Hamiltonian constraint.  In
three spacetime dimensions (3d), the length operator plays a role
analogous to the area operator in 4d \cite{ponzano}. For the 3d
Euclidean case, the eigenvalues of the length are discrete
\cite{ponzano}.  In this letter we point out that, surprisingly,
the spectrum of the length appears to become continuous in the
Lorentzian case.

This difference can be traced to the fact that in the usual tetrad/triad
formulation the canonical structure of general relativity is quite
different in 4d or in 3d.  In 4d the Lagrangian internal gauge group is
$SO(4)$ in the Euclidean case and the Lorentz group $SO(3,1)$ in the
Lorentzian case.  However, both these groups are reduced to a $SO(3)$
subgroup when going to the canonical theory, in order to solve the second
class constraints.  (An alternative approach where the internal Lorentz
group is not broken has been recently explored in
\cite{sergei,sergei&dima,richard}). In loop
quantum gravity \cite{loop}, the area operator turns out to be given by
the Casimir of the internal gauge algebra.  The Casimir of the $so(3)$
algebra has discrete eigenvalues, yielding a discrete spectrum for the
area.  In 3d, on the contrary, the internal gauge group of the canonical
theory is the same as in the Lagrangian theory: it is $SO(3)$ in the
Euclidean case and $SO(2,1)$ in the Lorentzian case. In 3d it is the
length operator that turns out to be given by the Casimir of the internal
gauge algebra.  (In this sense length in 3d is analogous to area in 4d.)
In the Euclidean case, the Casimir of $so(3)$ has discrete eigenvalues,
yielding discrete length.  But in the Lorentzian case, the Casimir of the
$so(2,1)$ algebra has discrete as well as continuous eigenvalues.  The
Casimir has opposite sign in the two cases, and a careful tracking of the
sign leads to the surprising result that spacelike intervals are not
quantized, while timelike intervals are.  This is contrary to the
expectation that discreteness is a feature of geometrical quantities at
fixed time.

Intuitively, one can visualize the geometry of the situation as
follows.  In 3d there is one timelike direction and two spacelike
directions.  In the Lorentz algebra there is --correspondingly-- one
rotation and two boosts.  The timelike direction is naturally
associated with the rotation.  In turn, the rotation (as opposite to
the boosts) is associated with the discrete spectrum.  The
timelike/spacelike character of the $so(2,1)$ unitary representations
was also emphasized by Witten in \cite{orbit}.

An analogous exchange (spacelike $\leftrightarrow$ continuous and timelike
$\leftrightarrow$ discrete instead of the contrary) was observed by
Barrett and Crane in \cite{BC} in a covariant spin foam
treatment of 4d quantum gravity, as well as in \cite{alex} in a
similar context.
In 3d, the same kind phenomenon was observed
by 't Hooft in the context of the canonical quantization of a
tesselated (Cauchy) surface \cite{hooft} (although it is also argued that
some imaginary part of space-like distances should be discrete)
and then again in the context of covariant spin foam models \cite{lpr}.  One
might then have suspected that this exchange is a feature of the spin
foam approach, in contrast with loop canonical results.  The result in
this paper rules out this idea, and shows that in 3d there is a
remarkable consistency between the results of the spin foam approach
(in \cite{lpr,davids}) and of the loop approach (here). This is
similar to results obtained through a covariant treatment of the canonical
theory in 4d \cite{sergei&dima,richard,larea},
where a continuous spectrum for spacelike
intervals was derived in contrast with the usual result obtained in
Loop Quantum Gravity. On the other hand, it is in contrast with the
compact group approach to 2+1 loop gravity, as developped in
\cite{jacek}, which uses a Wick rotation and derives a discrete
spectrum for space-like distances.

\section{$2+1$ Loop quantum gravity}

\subs{Canonical structure}

In 3d, general relativity can be formulated as follows.  The
gravitational field is represented by an $SO(2,1)$ connection
$A^i_\mu(x)$ and a triad $e^i_\mu(x)$.  Here $\mu=0,1,2$ is a
space-time (co-)tangent index, and $i=0,1,2$ is an internal index,
labelling a basis in the $so(2,1)$ algebra.  We will be working in a
space-time of signature $(-++)$, so that we raise and lower internal
indices using the flat metric $\eta_{ij}={\rm diag}[-++]$.  The action
is then given in terms of the triad and the field strength
$F^{i}_{\mu\nu}= \partial_\mu A^{i}_{\nu} -\partial_\nu A^{i}_{\mu}
+\eta^{ij}\epsilon_{jkl} A^{k}_{\mu} A^{l}_{\nu}$ ($\epsilon_{ijk}$
is the completely antisymmetric object) by
\begin{equation}
    S(A,e)
=\f{1}{G}\int \mathrm{tr}( e\wedge F)
=\frac{1}{G}\int d^3x \ \eta_{ij}\ \epsilon^{abc}\ e^i_a\
    F^{j}_{bc},
\end{equation}
where $G$ is a (rescaled) Newton constant.  We can perform the usual
Hamiltonian analysis, by choosing $x^0$ as the time evolution parameter
and $x^a=(x^1,x^2)$, as coordinates of the initial surface $\Sigma$, which
we take closed and orientable.
Then the action can be decomposed as
\beqs
    S&=& \frac{1}{G}\int dt \int_\Sigma dx^a \ \left( \eta_{ij}\,
    \epsilon^{ab}\ e^i_a\left( \partial_0 A^j_{b} -\partial_b A^j_{0}
    +\eta^{jk}\epsilon_{klm} A^l_{b} A^m_{0}\right) +
    \eta_{ij}\epsilon^{ab} e^i_0\ F^j_{ab} \right)\nonumber \\
&=& \f{1}{G}\int dt \int_\Sigma dx^a \left(
\epsilon^{ab}\eta_{ij} e^i_a\partial_0 A^j_{b}+
A_0^i\epsilon^{ab}(\eta_{ij}\partial_be_a^j+
\epsilon_{ijk}e^j_aA^k_b)
+e^i_0\eta_{ij}\epsilon^{ab}  F^j_{ab}
\right),
    \label{eq:deco}
\eeqs
where $\epsilon^{ab}=\epsilon^{0ab}$.  From this expression we can
read out that the canonical variables are $A^i_{a}(x)$, and their
conjugate momenta are $\pi^{a}_{i}(x) = \frac{1}{G} \eta_{ij}
\epsilon^{ab} e^j_b(x)$.  The fundamental Poisson bracket is therefore
\begin{equation}
    \{ A^i_a(x), e^j_b(y)\}=G\ \epsilon_{ab}\, \eta^{ij}\,
    \delta^{(2)}(x,y).
    \label{eq:pb}
\end{equation}
The Lagrange multipliers $A_0^i$ and $e^i_0$  enforce
the constraints
$\epsilon^{ab}{\cal D}_a e^i_b=0$ and $F^i_{ab}=0$, respectively.
The first one --the {\it Gauss law}--
generates the $SO(2,1)$ gauge transformations.
The second one forces the curvature to be flat. It
generates a variation of the frame field:
\beq
\left|
\begin{array}{ccl}
\delta e^i_a &=& {\cal D}_a \lambda^i \\
\delta A^i_a &=& 0.
\end{array}
\right.
\eeq
When the frame field is non-degenerate,
the second constraint can
be decomposed in a vector constraint
imposing invariance under 2d space diffeomorphism and a scalar constraint
(or Hamiltonian constraint) \cite{thiemann3d}.
More precisely, let us introduce the (co-)frame
field conjugate to the connection:
\beq
E^a_i=\epsilon^{ab}\eta_{ij}e^j_b,
\eeq
and the normal density vector
\beq
E^i=\f{1}{2}\epsilon^{ijk}\epsilon_{ab}E^a_jE^b_k.
\label{ndv}
\eeq
Using the facts that $\eta_{ij}E^iE^j=-\textrm{det}({}^2g)$
is the determinant of the 2-metric (the minus sign is due to the
Lorentzian signature of the metric $\eta$)
and that $E^a_iE^i=0$, we can decompose the flatness constraint as follows
\cite{thiemann3d}
\beq
N^iF^i=N^aV_a+NH.
\eeq
$N^a$ and $N$ are respectively the Shift and the Lapse, defined as:
\beq
N^i=N^a\epsilon_{ab}E^b_i+N\f{E^i}{\sqrt{\textrm{det}({}^2g)}}
\qquad\Leftrightarrow\qquad
N^a=\epsilon_{ijk}\f{E^iE^a_j}{\textrm{det}({}^2g)}N^k
\textrm{ and }
N=\f{N^iE^i}{\sqrt{\textrm{det}({}^2g)}}.
\eeq
$V_a$ is a vector constraint enforcing the invariance
under space diffeomorphisms
and $H$ is the Hamiltonian constraint:
\beq
\left\{
\begin{array}{ccc}
V_a &=&F^i_{ab}E^b_i \\
H &=&\f{1}{2}\epsilon_{ijk}F^i_{ab}\f{E^a_jE^b_k}{\sqrt{\textrm{det}({}^2g)}}
\end{array}
\right.
\eeq
In the present paper, we study the kinematical
structure of the quantum theory, which describes the
quantum geometry. We shall not deal with the dynamics.

\subs{Loop Quantization}

\label{loop}

A quantum theory is defined by a space of quantum states
and an algebra of operators.
In loop quantum gravity, the (kinematical) states of the geometry
are chosen to be cylindrical functions of the connection $A$.
A cylindrical function is
determined by an oriented graph $\Gamma$ (with $E$ edges)
and a function $\psi$ on $(SO(2,1))^E$. It is defined as
\beq
\Psi_{\Gamma,\psi}(A)=\psi(U_{1}(A),\ldots,U_E(A))
\eeq
where
\beq
U_{e}(A)={\cal P}\exp\left({\int_e ds \,\dot{\alpha}^a(s)\,
A^i_a(\alpha(s))\ \tau_i}\right).
\eeq
is the holonomy of the connection $A$ along the edge $e$ of the graph.
Here $\tau_i$ are the three generators of the $so(2,1)$
algebra.

Two basic operators are defined on the space of these functionals.
The first is the holonomy
of the connection $A$ along any loop. It acts multiplicatively on
the functionals of $A$.  The second is the operator value distribution
corresponding the field $e^i_a(x)$. This is given by the differential
operator
\begin{equation}
e^i_a(x) =
-i\hbar G\epsilon_{ab}\eta^{ij}\f{\delta}{\delta A^j_b(x)},
\label{op:e}
\end{equation}
where $\hbar G$ is the Planck length $l_P$ in three dimensions. The
quantum algebra of these operators provides a quantization of their
classical Poisson algebra.

Recall the identity \cite{jerzy}
\begin{equation}
\f{\delta}{\delta A^j_b(x)}U_e(A)=
\int_e ds \, \frac{de^b(s)}{ds} \delta^{(2)}(e(s),x)\
U_{e_1(s)}(A) [X^i U_{e_2(s)}](A),
\end{equation}
where $e_1(s)$ and $e_2(s)$ are the two parts in which $e$ is
split by the point $x$, and $X^i$ is the generator of the left
action of $SO(2,1)$ on the functions on the group.
Using this, we have immediately
the action of the triad field operator density on the cylindrical states.
If $x$ is not on $\Gamma$, this action vanishes. Assuming for simplicity that $x$
is in the interior of the edge $e$, it is
\begin{equation}
e^i_a(x) \Psi_{\Gamma,\psi}=
-i\hbar G\epsilon_{ab}\eta^{ij}
\int_e ds \ \frac{de^b(s)}{ds} \delta^{(2)}(e(s),x)\
\Psi_{\Gamma_s,X^i\psi_s}
\label{triadop2}
\end{equation}
where $\Gamma_s$ is the graph $\Gamma$ with $e$ split into $e_1(s)$ and $e_2(s)$,
\begin{equation}
\psi_s(U_1, ..., U_{e_1(s)},U_{e_2(s)}, ..., U_E)=
\psi(U_1, ..., U_{e_1(s)}U_{e_2(s)}, ..., U_E)
\end{equation}
and $X^i$ acts on the variable $U_{e_2(s)}$ by left multiplication.

The Gauss constraint imposes the states to be invariant
under $SO(2,1)$ gauge transformation
of the connection. This implies that the functions
$\psi$ must be invariant at the nodes in the following sense
\beq
\psi(U_1,\dots,U_E)=
\psi(k_{s(1)}U_1k_{t(1)}^{-1},\dots,k_{s(E)}U_Ek_{t(E)}^{-1}),
\hspace{3em}
\forall k_v \in SU(1,1),
\label{gaugec}
\eeq
where $s(e)$ is the source node of the edge $e$ and $t(e)$ its
target node.

The scalar product on the space of these states is determined by the
requirement that real classical quantities be represented by
hermitian operators. Observe first that
any cylindrical functional $\Psi_{\Gamma\psi}$ determined by a graph $\Gamma$
can be rewritten as a cylindrical functional determined by a graph $\Gamma'$
that contains $\Gamma$ (such that $\Gamma$ is a subgraph of $\Gamma'$)
\begin{equation}
\Psi_{\Gamma\psi}(A)=\Psi_{\Gamma'\psi'}(A);
\end{equation}
indeed, it is sufficient to take $\psi'$ as
independent from the edges of $\Gamma'$ that are not in $\Gamma$. Using this, it
is clear that we can always write any two cylindrical functionals in terms of the
same graph. Using this fact, in the cases in which the gauge group is compact
the scalar product between two states is defined by
\begin{equation}
(\Psi_{\Gamma\psi},\Psi_{\Gamma\psi'})
=\int dU_1 ... dU_E \ \ \overline{\psi(U_1...U_E)}\ \psi'(U_1...U_E).
\label{measure}
\end{equation}
The holonomy operator acts by multiplication
and the reality condition for the connection
$\what{A^i_a}^\dagger=\what{A^i_a}$ is trivially implemented.
The hermicity of the frame field operator,
implies simply that the operator $iX^i$ in (\ref{triadop2}) be hermitian,
namely that the  measure $dU_e$ is invariant under the action of the group.
That is, it must be the Haar measure. Notice that if the states are
independent from a link $e$ of the graph, the integration in $dU_e$
becomes irrelevant thanks to the normalization of the Haar measure
of a compact group: $\int dU_e =1$.

In our case, however, $SO(2,1)$ is non compact. This implies that more
care is required in the definition of the scalar product. A gauge invariant state,
in particular, is constant along an orbit of the group, and the
integral in (\ref{measure}) diverges along this orbit.  Similarly, a moment
of reflection shows that the  triad operator may send a finite norm state
into an infinite norm state.  These divergences can be taken care
of by restricting the integration in (\ref{measure}) to a suitably
chosen subset of integration variables, such that the spurious integrations
along gauge orbits are eliminated. The construction amounts to
divide out the volume of the gauge group. In  \cite{spinnet} it was shown 
that this can be done systematically. 

More precisely,
we call a {\it connection} on a graph $\Gamma$ the assignment of group
elements $U_e$ to each link of the graph and we define
$G_{\Gamma}=G^{\otimes E}/G^{\otimes V}$ (where $\Gamma$ has $E$ edges
and $V$ vertices)
the space of the equivalence classes of these connections under the gauge 
transformation (\ref{gaugec}). Notice that a function $\psi$ satisfying
(\ref{gaugec}) is a function on $G_{\Gamma}$. 
It is shown in 
\cite{spinnet} that the Haar measure naturally defines a measure
$d\mu_{\Gamma}$ on $G_{\Gamma}$.  Each graph has then an associated
Hilbert space $L^{2}(G_{\Gamma},d\mu_{\Gamma})$, and we can 
replace (\ref{measure}) by the scalar product in this Hilbert space. 
Holonomy functionals of the connection with support on $\Gamma$ 
act by multiplication on this space and this implements the reality 
condition for the connection.
It is then proven in \cite{spinnet} that any gauge invariant 
operator constructed with the triad field is well defined 
$L^{2}(G_{\Gamma},d\mu_{\Gamma})$ and is hermitian in this measure,
which implements the hermicity condition for the frame fields.
We refer the reader to \cite{spinnet} for all details\footnotemark.
\footnotetext{The tricky point (and the difference with the compact group case)
in the construction is that the space $L^2_\Gamma$
for a graph $\Gamma$ can not be embedded in the space $L^2_{\Gamma'}$
associated to a bigger graph. As a consequence, the full space of quantum
states of geometry, obtained by gluing these $L^2_\Gamma$ spaces
together (summing over graphs),
can not be obtained as a projective limit anymore (as in the
compact group case) and, therefore, doesn't seem to be a $L^2$
space. On the other hand, it is still a Hilbert space, with a 
structure similar to a Fock space.}

Using the Plancherel decomposition of $L^2$ functions on the group
(with  respect to the Haar measure), an orthonormal basis of states in
$L^2_\Gamma$ can be constructed as spin networks and involves the
infinite dimensional unitary representations of $SO(2,1)$.
Recall that $L^2$ functions over the group can be expanded over an
orthonormal basis provided by irreducible representations of the 
group. This is called the Plancherel decomposition of the $L^2$ 
functions on the group.  The representations appearing in this 
expansion are the ones of the {\it principal} and {\it discrete} 
series of {\it unitary} representations, and will be described
below in Section III.

Then, we can construct an orthonormal basis of gauge invariant states as follows.
Once the graph $\Gamma$ is fixed, we choose a principal unitary
$SO(2,1)$ irreducible representation ${\cal I}_e$ (entering the
Plancherel decomposition)
for each edge $e$ of  the graph.
Contract the representation matrices of the holonomies
$U_e^{{\cal I}_e}$ in these representations using a $SO(2,1)$ intertwiner 
at each node (intertwining the representations associated to the edges 
incident to the node). The resulting function is an $SO(2,1)$ spin network 
functional. It depends on the graph, the representations associated to the 
edges and the intertwiners associated to the nodes. The set of these 
functionals (for all graphs, all choices of irreducible representations and 
intertwiners out of a basis in the linear space of the intertwiners) form 
a complete orthonormal (generalized) basis of gauge invariant cylindrical 
functionals.

\subs{Length operator}

The length of a
differential curve $c:\tau\in [0,1]\rightarrow c(\tau)\in\Sigma$ is
given by
\begin{equation}
    L_c=\int_{[0,1]} d\tau\ \sqrt{\dot{c}^a(\tau)\, \dot{c}^b(\tau)\ g_{ab}(c(\tau))}
    =\int_{[0,1]} d\tau\ \sqrt{\dot{c}^a(\tau)\, \dot{c}^b(\tau)\ \eta_{ij}\,
    e^i_a(c(\tau))\, e^j_b(c(\tau))}.
    \label{length}
\end{equation}
For simplicity of the notations, let us introduce the vector
\beq
e^i(c(\tau))=e^i_a(c(\tau))\dot{c}^a(\tau).
\eeq
In the study of the length, we will restrict ourselves to the case
in which the norm $\eta_{ij}e^ie^j$ of the vector $e^i$
doesn't change sign along the curve.
That is, we require the curve to be entirely time-like or
entirely space-like.
Notice that in the case of a  time-like curve,  that is $\eta_{ij}e^ie^j<0$,
there is another gauge invariant quantity besides the norm of $e^i$:
the sign of $e^0$. This sign is
invariant under $SO(2,1)$ and registers the time orientation, past or future,
of the curve.

The length of a space-like curve ($\eta_{ij}e^ie^j>0$) is be defined
by \Ref{length} as a real number.
In the case of a time-like curve ($\eta_{ij}e^ie^j<0$)
it is convenient to define an oriented real time interval taking
into account the time orientation as
\beq
T_c=\textrm{sign}(e^0)\int_{[0,1]} dt\ \sqrt{|\eta_{ij}e^ie^j|}.
\label{time}
\eeq
The quantum operator representing the classical length is obtained
replacing the triad field $e^i_a(x)$ with the corresponding
quantum operator \Ref{op:e} in these expressions.
We now study the action of this length operator on spin network states, following
the example of the area operator in $3+1$ loop quantum gravity
\cite{primer,discrete2}. Our concern here is not in the details of the regularization
of this operator, which have been extensively discussed elsewhere, but just
on the particular features of the Lorentzian 2+1 case.

Consider a curve $c$ and a spin network state such that the
curve and the underlying graph intersect only once
and not at a node of the graph.  We consider the action of the length operator of the curve
$c$ on this state. (What follows can be easily generalized
to multiple intersections and to intersections at nodes.)
Call $\gamma$ the edge of the spin network intersected by the curve $c$. Let
${\cal I}$ be the irreducible representation associated to
$\gamma$.

The action of $e^i(x)$ on the spin network state inserts the
generator $X^i$ in the state. The action of  $X^i$ on the
representation ${\cal I}$ is given by the generator
$X^i_{\cal I}$ in this representation. We obtain easily
\begin{equation}
{L_c}\Psi^{({\cal I})}=\hbar G \left[\int_c d\tau\,\sqrt{
    \left(\int_\gamma ds\,\epsilon_{ab}\dot{c}^a(\tau)
    \frac{d\gamma^b}{ds}\delta^{(2)}(\gamma(s),c(\tau))
    \right)^2
    \left(-\eta_{jk}X^j_{({\cal I})}X^k_{({\cal I})}\right)
    }\ \right]\ \ \Psi^{({\cal I})},
    \label{eq:l}
\end{equation}
which shows that the spin networks are eigenvectors of the length operator.
The integral quantity
\beq
\int_c d\tau\,\int_\gamma ds\,\left|\epsilon_{ab}\dot{c}^a(\tau)
    \frac{d\gamma^b}{ds}\delta^{(2)}(\gamma(s),c(\tau))\right|
\eeq
is the number of intersections between the curve $c$
and the edge $\gamma$ of the spin network graph.
More precisely, $\epsilon_{ab}\dot{c}^a(\tau)\gamma'^b(s)$ is
the Jacobian of the transformation between orthonormal coordinates
$(x_1,x_2)$ and the local coordinates $(\tau,s)$. This works only when
the curve $c$ and the edge $\gamma$ are not tangential else the action
of ${L_c}$ vanishes.
Here we have assumed the intersection number to be 1,
so that the action of the length operator reduces to
\begin{equation}
{L_c}\ \Psi^{({\cal I})}= \hbar G\
    \sqrt{-\eta_{jk}X^j_{({\cal I})}X^k_{({\cal I})}}\ \ \Psi^{({\cal I})}= \hbar G\
    \sqrt{q^{({\cal I})}}\ \Psi^{({\cal I})}.
    \label{eq:eigenvalue}
\end{equation}
where $Q=-\eta_{ij}X^i X^j$
is the Casimir operator for $SO(2,1)$ and $q^{({\cal I})}$ is its
value in the representation $\cal I$. This gives
the length spectrum of $2+1$ gravity in its loop quantized version, which we
study in the next section.
Depending on the sign of $q^{(\cal I)}$, we get either a space-like length, a time-like length
or a null curve.
If the case of a time-like or null curve, we further need to specify the orientation
observable $\textrm{sign}(e^0)$. This should correspond to some data encoded
in the representation ${\cal I}$. More precisely, it should be the sign
of the operator $\what{e^0}$, which should thus have a spectrum
with only positive eigenvalues or only negative eigenvalues.

The length operator of a curve acting on a
spin network which it intersects
many times is given by a contribution for each edge intersected.
\beq
{L_c}\ \Psi=l_P\sum_{e|e\cap c\ne\emptyset}
\sqrt{q^{({\cal I}_e)}}\, \Psi.
\eeq

\section{Representations of $SO(2,1)$ and the length spectrum}

We are interested in the
group $SO(2,1)$, the Lorentz symmetry group of $2+1$ gravity.
Its Lie algebra is of dimension 3 and generated in its
fundamental representation by the following three matrices
\footnotemark
\footnotetext{In fact, this is the fundamental representation
for $SU(1,1)$, which is the double cover of $SO(2,1)$.
In the present paper, we will use only the group $SO(2,1)$ for its
representation theory is simpler. However, all the results can be
obviously extended to the case of $SU(1,1)$ and we present its
representation theory in appendix \ref{su11}. Let us nevertheless
point out that  $SU(1,1)$ is not the universal cover
of $SO(2,1)$ unlike the Euclidean case where $SU(2)$ was actually
the universal cover of $SO(3)$.}:
\begin{equation}
\tau_0=\frac{i}{2}\mat{cc}{1 & 0 \\ 0 & -1}, \quad
\tau_1=\frac{1}{2}\mat{cc}{0 & 1 \\ 1 & 0}, \quad
\tau_2=\frac{i}{2}\mat{cc}{0 & -1 \\ 1 & 0}.
\label{tau}
\end{equation}
The commutation relations between these generators  are:
\beq
[\tau_0,\tau_1]=-\tau_2 \qquad [\tau_1,\tau_2]=\tau_0 \qquad [\tau_2,\tau_0]=-\tau_1.
\eeq
One can check that these are the right signs for the symmetry group
$SO(2,1)$ of a $(-,+,+)$ Lorentz space.
Let $X_i$ be the generators of a linear representation of the group. They are
linear operators that satisfy
\beq
[X_0,X_1]=-X_2 \qquad [X_1,X_2]=X_0 \qquad [X_2,X_0]=-X_1.
\eeq
It is important not to confuse the hermicity properties of the
matrices $\tau_i$ and the hermicity properties of the $X_i$.
As we have discussed above, the representations playing a role
in quantum gravity are the ones appearing in
the Plancherel decompositions of the $L^2$ functions with respect
to the Haar measure. These representations are unitary i.e
the linear operators $iX_i$ are hermitian. Indeed, these are
(up to constants) precisely the quantities corresponding the triad
field operator, and their hermicity reflects the fact that the triad
field is real.

It is useful to study the algebra using the operators $H$ and $J_\pm$
defined as:
\beq
H=-iX_0 \qquad J_\pm = \pm X_1 +i X_2
\eeq
with the commutation relations:
\beq
[H,J_\pm]=\pm J_\pm \qquad [J_+,J_-]=-2H.
\label{comm}
\eeq
The difference with the real algebra $so(3)$ is the minus sign in the second
commutation relation. The Casimir operator is
\beq
Q=(X_0)^2-(X_1)^2-(X_2)^2
=-H^2+\frac{1}{2}(J_+J_-+J_-J_+).
\label{casimir}
\eeq
The reality conditions expressing that the $iX_i$ are hermitian are
expressed as:
\beq
H^\dagger=H \qquad
(J_\pm)^\dagger=J_\mp.
\label{reality}
\eeq

\subs{Representations of $SO(2,1)$}

The representations of $SO(2,1)$ can be studied in the same type
of basis as for $SO(3)$. Indeed it is easy to check that
\begin{equation}\label{rep}
    \left\{
    \begin{array}{ccc}
    H|m\ra&=& m|m\ra, \\
    J_+|m\ra&=& (q +m(m+1))^{1/2}|m+1\ra, \\
    J_-|m\ra&=& (q +m(m-1))^{1/2}|m-1\ra
    \end{array}
    \right.
\end{equation}
gives a representation of $SO(2,1)$ on the space spanned by the vectors
$\{|m\ra\ , \ m\in{\mathbf Z}\}$. The the parameter $q$ gives
the value of the Casimir operator. \footnotemark
\footnotetext{If we replace $m$ by
$m+1/2$ everywhere in (\ref{rep}) we get a representation of $SU(1,1)$. If we
replace $m$ by $m+\alpha$, $0<\alpha<1$ we get a representation of the
universal cover of $SO(2,1)$}

The unitary representations are infinite dimensional since $SO(2,1)$
is non-compact. Their Casimir operator is
hermitian $Q^\dagger=Q$. This implies that  $q$ is real.
Let us consider the different representations obtained for real
values of the parameter $q$.

Consider first the case of a negative Casimir $q\le0$.
For generic values,
the representation obtained is irreducible. However
$(q+m(m+1))$ can take some negative values, and this contradicts
the unitarity relation $(J_\pm)^\dagger=J_\mp$.

For the special values $q=-n(n-1)\le0$, with $n\in \N^*$,
$(q+m(m+1))$ vanishes for values $m=n-1$ and $m=-n$. Therefore,
the representation is not irreducible. In facts, it decomposes
into 3 representations.

There are ``intermediate'' representations, called $V^n$,
which are finite dimensional. They are
spanned by the vectors $\{|m\ra\ , \ -(n-1)\le m\le (n-1)\}$.
They are  the same representation
as the finite irreducible (spin $j=n-1$) representation of $SO(3)$.
However $(q+m(m+1))<0$ and we have $(J_\pm)^\dagger=-J_\mp$, which violates
the reality conditions: they are not unitary.

The other two representation are infinite dimensional.
The upper one ${\cal D}^+_n$ is a lowest weight representation spanned by values
$m\in n+\N$. The lower one ${\cal D}^-_n$ is a highest weight representation
spanned by $m\in -(n+\N)$. These representations are unitary.

For a positive value of the Casimir $q>0$,
$(q+m(m+1))=q-1/4 +(m+ 1/2)^{2}¥$ stays always positive and we get infinite dimensional
unitary representations spanned by all $m\in{\mathbf Z}$.
$0<q<1/4$ labels the exceptional series whereas
$q>1/4$ labels the principal series. The representations of the principal
series are denoted ${\cal C}_s$, with $q=s^2+1/4$.

The unitary irreducible representations ${\cal D}^+_n$, ${\cal D}^-_n$ and
${\cal C}_s$ are the ones coming into the {\it Plancherel
decomposition} of a $L^2$ function $f$ on the group $SO(2,1)$:
\beqs
f(g)&=&
\sum_{n\ge 1}\ (2n-1)\ \textrm{Tr}(f^+_nR_{{\cal D}^+_n}(g))
+\sum_{n\ge 1}\ (2n-1)\ \textrm{Tr}(f^-_nR_{{\cal D}^-_n}(g))
\nonumber \\
&&+\int_{s>0} ds\ \frac{\coth(\pi s)}{4\pi s}\
\textrm{Tr}(f_sR_{{\cal C}_s}(g)),
\eeqs
Notice that the continuous
representations start at $q=1/4$ instead of $q=0$.

\subs{Length spectrum}

The eigenvalues of the length operator associated to a curve
are given by the square root of the values
of the Casimir operator of the representation carried by the edge
that the curve intersects. A continuous representation
${\cal C}_s$ has {\em positive} Casimir and correspond to a space-like length with eigenvalue
\beq
L_s=\sqrt{s^2+1/4}.
\eeq
Notices the gap $1/2$. It implies that there exists a minimal space-like
length even if the spectrum is {\it continuous}.

A discrete representations ${\cal D}^\epsilon_n$
($\epsilon=\pm$ and $n\in\N$)
has {\em negative} Casimir and
corresponds to a time-like curve.
Its past or future orientation $\textrm{sgn}(e^0)$
is given by $\epsilon$. Indeed, $\epsilon$ is the sign of the
(eigenvalues of the) generator $H$, which is the operator
quantizing $e^0$. Then the length spectrum corresponding to
a time-like curve will be {\it discrete}. The eigenvalues of the
observable (\ref{time}) are
\beq
T_{\epsilon,n}=\epsilon\sqrt{n(n-1)}.
\eeq
Notice that the eigenvalues do not have  equal spacing.
Notice also that the first discrete representations
${\cal D}^\pm_{n=1}$ have vanishing Casimir and length eigenvalue,
and thus correspond to a null curve, the sign $\pm$ still corresponding
to the past or future orientation of the curve. See Figure 1.

More precisely, the eigenvalues of the length operators are given by
any sum of these eigenvalues. Each term of the sum corresponds to one
intersection between the graph of the state and the curve $c$.
Note that the gap in the real axis between $0$ and $1/2$ correspond to a class of unitary
 representations. However, these representations
are not $L^2$ and have a vanishing Plancherel measure (they do not
come in the Plancherel decomposition): they are called the
{\it complementary series} of representations.

\begin{figure}[t]
\begin{center}
\psfrag{p1}{$1^+$}
\psfrag{m1}{$1^-$}
\psfrag{nm2}{$n=2^-$}
\psfrag{np2}{$n=2^+$}
\psfrag{np3}{$n=3^+$}
\psfrag{np4}{$n=4^+$}
\psfrag{r2}{$T=i\sqrt{2}\sim 1.41$}
\psfrag{r6}{$T=i\sqrt{6}\sim 2.45$}
\psfrag{r12}{$T=i\sqrt{12}\sim 3.46$}
\psfrag{s0}{$s=0,L=\f{1}{2}$}
\includegraphics[width=6.5cm]{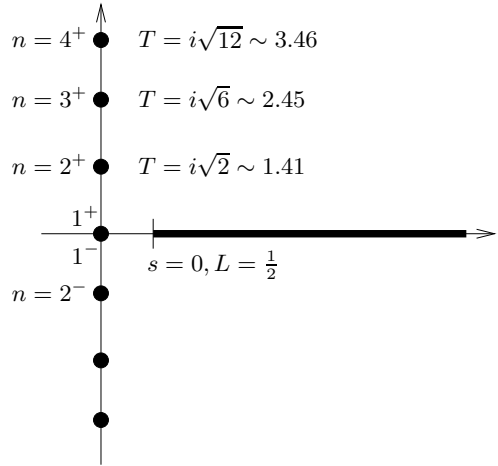}
\end{center}
\caption{The spectrum of the length operator.}
\end{figure}

\subs{Variants}

Alekseev and al. in the context of loop quantum gravity \cite{alekseev}
(for the groups $SO(3)$ and $SU(2)$) and Freidel-Krasnov
in the context of spin foams \cite{sf} have given some arguments
for a possible correction to the above
length spectrum. We can use the symmetric quantization map for
quantizing $\what{e^i}$ as the derivation on the Lie algebra (as a
vector space) instead of the derivation on the Lie group.\footnotemark
\footnotetext{As an example, consider the following analogy
with the model of the free particle
over the group manifold \cite{spinnet}. The theory is
defined by the action
$$
S=\f{1}{2}\int dt\, \textrm{Tr}((g^{-1}\pp_tg)^2).
$$
As conjugate variable for the configuration variable $g$, we
can choose
either the canonical momentum $p=\pp_tg^{-1}$, which is commutative, or
the non-commutative Noether charge $\Pi=g\pp_tg^{-1}$. The last generates
the (right) group multiplication and
satisfies $\{\Pi_X,\Pi_Y\}=\Pi_{[X,Y]}$,
where $\Pi_X=\textrm{Tr}(X\Pi)$ is the $X$ component of $\Pi$.
}
Then the Casimir operator $q$ gets shifted to $q-1/4$
\footnotemark
\footnotetext{See \cite{spinnet} for details on $SO(2,1)$ and $SU(1,1)$ and explicit
expressions for the Laplacian and the characters.}
and the length spectrum becomes
\beq
\left|
\begin{array}{ccccc}
L_s&=&s & \textrm{for space-like } &{\cal C}_{s>0}\\
T_{\epsilon,n}&=&\epsilon\left(n-\f{1}{2}\right) & \textrm{for time-like }&
{\cal D}^{\epsilon=\pm}_{n\ge1}
\end{array}
\right.
\label{lengthspectrum}
\eeq

There is now a minimal time-like interval and no null representation
given by the discrete series. Also, the time-like length spectrum becomes
equally spaced. On the other hand, the initial
gap for the space-like lengths disappear and there is the possibility
of a null curve in the limit $s\rightarrow 0$.
This second length spectrum
fits better with the algebraic data and
with the Lorentzian 3d Spin Foam picture \cite{lpr,davids}, see also
\cite{louapref} for the asymptotics of 6-j symbols.
See Figure 2.
In this version of the length spectrum,
null representations are also present in the representation
theory of $su(1,1)$. There are two extra discrete
representations given by $n=1/2^\pm$ (present only for the group
$SU(1,1)$ and not for $SO(2,1)$); these representations are unitary
but {\it not} $L^2$, they are called the {\it limit of discrete series}.
\begin{figure}[t]
\begin{center}
\psfrag{pm0}{$(1/2^\pm)$}
\psfrag{p1}{$1^+$}
\psfrag{m1}{$1^-$}
\psfrag{nm2}{$n=2^-$}
\psfrag{np2}{$n=2^+$}
\psfrag{np3}{$n=3^+$}
\psfrag{l1}{$L=i\f{1}{2}$}
\psfrag{l2}{$L=i\f{3}{2}$}
\psfrag{l3}{$L=i\f{5}{2}$}
\includegraphics[width=6.5cm]{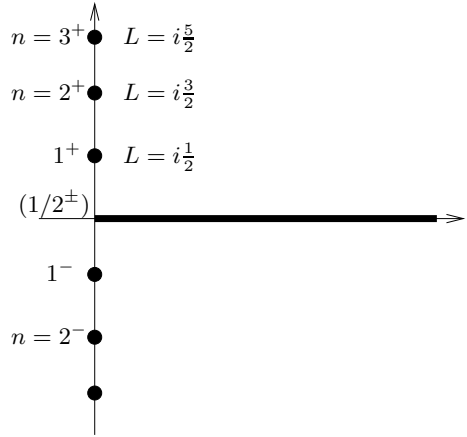}
\caption{The spectrum of the symmetric length operator.}
\end{center}
\end{figure}


\section{Area spectrum}

\subs{Area operator}

The area of a surface $\S$ embedded in the canonical (closed  and orientable)
surface $\Sigma$ is given by
\beq
\A_\S=\int_\S ds^2 \sqrt{\textrm{det}({}^2g)}
\eeq
where $g_{ab}=e^i_ae^j_b\eta_{ij}$ is the 2-metric on $\Sigma$.
We now study the quantum operator corresponding to this quantity.

The determinant of the metric can be written as
$\textrm{det}({}^2g)=-\eta^{ij}E_iE_{j}$
in term of the normal density  vector $E_i(x)=\f{1}{2}\epsilon_{ijk}\epsilon^{ab}e_a^j(x)e_b^k(x)$
introduced in (\ref{ndv}).
When acting on a spin network, the frame field operator
has a non-vanishing action only
if $x$ belongs to the graph. When $x$ is in the middle
of an edge, the action of the operator is proportional to
the tangent $\dot{\gamma}_a(s)X^i$ of the edge
(see \Ref{triadop2}).
It follows immediately that $\epsilon^{ab}\dot{\gamma}_a\dot{\gamma}_b=0$,
vanishes and therefore
the operator corresponding to $E_i(x)$ gives zero.
The only points at which $E_i(x)$ will have a non-vanishing
action are the nodes of the graph: {\it the area operator
has contributions only from the nodes}.

To compute the action of the area operator of a surface $\S$ on a spin network state,
we cut up the surface into small bits,
each containing at most one node of the spin network. We can thus
restrict ourselves to the study of a (elementary) surface containing only one
node $n$ of the spin network on which the area operator acts.
For simplicity, we also restrict ourselves to the case  where
the node is 3-valent.
To define the area operator, we need to choose an {\it orientation} for
$\Sigma$, even though the final result will be independent of
the chosen orientation. This corresponds to choosing a consistent
{\it ordering} of the three edges incident on each node of the graph.

The node $n$ has three incident edges
$e=1,2,3$ (following the orientation) with $SO(2,1)$ representation
${\cal I}_e$.
To begin with, define an auxiliary operator acting at the node $n$ by
the insertion of some $X$ operators
\beq
 {\wtl{E_i}} \Psi^{{\cal I}_1{\cal I}_2{\cal I}_3}=
 -\frac{l_P^2}{2} \epsilon_{ee'}\epsilon_{ijk}
X^j_{{\cal I}_{e}}X^k_{{\cal I}_{e'}}
\Psi^{{\cal I}_1{\cal I}_2{\cal I}_3},
\eeq
where $e,e'$ are any two edges meeting at $v$ and
$\epsilon_{ee'}$ registers the orientation of the two
edges around the node.
Using the fact  that
$\vec{X}_{{\cal I}_1}+\vec{X}_{{\cal I}_2}+\vec{X}_{{\cal I}_3}=0$,
one can get a more symmetric expression for ${\wtl{E_i}}$
by summing the above expression over the couples of edges
$(e,e')$, with a symmetry factor $1/3$.

We now look at the action of $\what{E_i}(x)$ at the node $n$, a
direct computation using (\ref{triadop2}) gives .
\beq
\what{E_i}(x) \Psi_{v}^{{\cal I}_1{\cal I}_2{\cal I}_3}=
\alpha(x,v)\ {\wtl{E_i}} \Psi_{v}^{{\cal I}_1{\cal I}_2{\cal I}_3},
\eeq
where the geometrical factor is
\beq
\alpha(x,v)=
\sum_{e,e'}\ \int ds dt\ \delta^{2}(x,\gamma_{e}(s))\
\delta^{2}(x,\gamma_{e'}(t))\
|\epsilon_{ab} \gamma^a_{e}(s) \gamma^b_{e'}(t)|.
\eeq
We can regulate this factor and see that it is just proportional to
$\delta^{2}(x,v)$.
Therefore the area operator acts on the spin network as:
\beqs
\A_\S\,\Psi^{{\cal I}_1{\cal I}_2{\cal I}_3} &=&
\sqrt{-\eta^{ii'}{\wtl{E_i}}{\wtl{E_i'}}}
\,\Psi^{{\cal I}_1{\cal I}_2{\cal I}_3} \\
&=&
l_P^2
\sqrt{-\f{1}{4}
\eta^{ii'}\epsilon_{ijk}\epsilon_{i'j'k'}
X^j_{{\cal I}_1}X^k_{{\cal I}_2}
X^{j'}_{{\cal I}_1}X^{k'}_{{\cal I}_2}}
\ \Psi^{{\cal I}_1{\cal I}_2{\cal I}_3} \nonumber\\
&=&
l_P^2\f{1}{2}
\sqrt{\left(
(\vec{X}_{{\cal I}_1})^2(\vec{X}_{{\cal I}_2})^2
-(\vec{X}_{{\cal I}_1}.\vec{X}_{{\cal I}_2})^2
\right)}
\Psi^{{\cal I}_1{\cal I}_2{\cal I}_3} \nonumber \\
&=&
l_P^2\f{1}{2}\sqrt{
|\vec{X}_{{\cal I}_1}\wedge\vec{X}_{{\cal I}_2}|^2
}
\Psi^{{\cal I}_1{\cal I}_2{\cal I}_3}.
\eeqs
Using the fact  that
$\vec{X}_{{\cal I}_1}+\vec{X}_{{\cal I}_2}+\vec{X}_{{\cal I}_3}=0$,
we can express the above factor in terms of the Casimir operators
$q^{{\cal I}_\alpha}$
of the three representations. Therefore the area operator is
diagonal in the spin network basis with eigenvalue:
\beq
\A_\S =
l_P^2
\f{1}{2}\sqrt{
q^{{\cal I}_1}q^{{\cal I}_2}-
\f{1}{4}(q^{{\cal I}_3}-q^{{\cal I}_1}-q^{{\cal I}_2})^2
}.
\eeq
This fits with the definition of the area of a geometrical triangle
defined by the length of its three edges given by
$L_\alpha=\sqrt{q^{{\cal I}_\alpha}}$.
More precisely, the above
formula can be rewritten as
\beq
\A_\S =
l_P^2 \f{1}{4}
\sqrt{
(L_1+L_2+L_3)
(-L_1+L_2+L_3)
(L_1-L_2+L_3)
(L_1+L_2-L_3)
}.
\eeq
This is consistent with the results on the length operator since
$L_\alpha=\sqrt{q^{{\cal I}_\alpha}}$ is precisely the length spectrum
associated to the spin network edge $\alpha$.

In conclusion, a labelled spin network has a geometrical interpretation as
a two-dimensional discrete triangulated
manifold. The faces have finite area and are dual to the nodes of the graph.
Faces are separated by edges with finite length, dual to the link of the graph.

\subs{$SO(2,1)$ Intertwiners and Area spectrum}

To find the spectrum of the area operator explicitly,
we need to characterize the admissible nodes i.e the
possible triplets of representations
$({\cal I}_1,{\cal I}_2,{\cal I}_3)$. This corresponds to the
existence of an intertwiner between these three representations.
Consider the decomposition of the tensor product of two
$SO(2,1)$ representations.
Not all kinds of principal representations show up in the decomposition of the
tensor product of two principal representations.   We have (see \cite{davids}, where
explicit expressions of the corresponding Clebsch-Gordon coefficient are
given)
\begin{equation}
    {\cal D}^\pm_{n_1}\otimes{\cal D}^\pm_{n_2}=
    \bigoplus_{n\ge n_1+n_2}{\cal D}^\pm_n,
\label{n+}
\end{equation}
\begin{equation}
    {\cal D}^+_{n_1}\otimes{\cal D}^-_{n_2}=
    \bigoplus_{n=1}^{n_1-n_2}{\cal D}^+_{n} \oplus
    \bigoplus_{n=1}^{n_2-n_1}{\cal D}^-_{n} \oplus
    \int ds \ {\cal C}_s,
\end{equation}
\begin{equation}
    {\cal D}^\pm_{n_1} \otimes {\cal C}_{s_2} =
    \bigoplus_{n\ge 1} {\cal D}^\pm_{n} \oplus
    \int ds \ {\cal C}_s,
\end{equation}
\begin{equation}
    {\cal C}_{s_1} \otimes {\cal C}_{s_2}=
    \bigoplus_{n\ge 1} {\cal D}^+_{n} \oplus
    \bigoplus_{n\ge 1} {\cal D}^-_{n} \oplus
    2\int ds\  {\cal C}_s.
\label{ccc}
\end{equation}
Precisely as in the Lorentzian 3d spin foam model \cite{davids,lpr},
these decomposition rules can be interpreted as describing the relations
between equivalence classes of
triangles (under the action of $SO(2,1)$) in Minkovski 3d
space. Equivalently, they can be interpreted as sum rules of 3-vectors.
More precisely, we can associate
the ${\cal D}^\pm_n$ representations to future or past
oriented time-like vectors with norm $L_n=n-1/2$ and the ${\cal C}_s$ to
space-like vectors with norm $L_s=s$. Notice that this is precisely the
association emerged from the spectrum of the length operator.
Then, equation \Ref{n+} corresponds to the fact that
summing two time-like vectors (with the
same orientation) gives a time-like vector of the same orientation.
Furthermore, the sum rule respects the anti-triangular
inequality $L_n\ge L_{n_1}+L_{n_2}$.
Similarly, equation \Ref{ccc}
corresponds to summing
triangles formed with  space-like edges. The result can be
space-like or time-like and
there is no (anti-)triangular inequality.
\footnotemark
\footnotetext{Notice that this implies that
restricting the theory to the continuous series of representations
in order to deal with solely space-like surfaces does not work,
unless one also imposes by hand a triangular inequality between
the continuous representations, which does not seem very natural from the
point of view of representation theory.}.

Finally, the area operator eigenvalues give precisely the area of the
different types of triangles obtained by summing two vectors as
described by these tensor product decomposition rules.

\section{Considerations}

\begin{itemize}

\item Our result is not definitive since we have considered only the
kinematics of the theory, and not its dynamics. It is not unconceivable
that the dynamics could constraint the representation of the operator
algebra in some unexpected way. Furthermore, questions remain open on
the definition of the full Hilbert space of the theory for non compact
groups \cite{spinnet}.

\item One may wonder how the length operator can have eigenvalues that
correspond to both signatures.  Since we use the canonical formalism,
the curve $c$ lives on the initial value surface.  If this is
spacelike, how can the curve be timelike?  The answer is the following.
In the canonical formalism considered, we have never imposed the
condition that the metric be spacelike on the initial surface.  In
fact, the canonical formalism is rather
flexible in this regard.  In 4d, one usually breaks down the Lorentz
group to a three dimensional rotation group.  In doing so, one gauge
fixes certain components of the tetrad to fixed values (with a well
defined sign), and this forces the remaining components, which form
the triad, to be spacelike.  Nothing similar happens in the canonical
formulation of the 3d theory considered here. Therefore, unless
one explicitly imposes so, the initial value surface has no
determined signature.

\item We recall that the length, as the area in 4d, is not a gauge
invariant operator, and its quantization has to be properly
interpreted as an indication of the corresponding quantization of a
suitable quantity defined intrinsically by the dynamical variables
themselves, as physical geometrical quantities measured in the
laboratory always are.  In general, the simplest way to do so is to
couple dynamical matter to the gravitational field and use this matter
as a physical reference frame \cite{matter,matter2}.  This also
explains how the rich structure given by the length operators can be
read out from the relatively simple 3d theory, which is topological,
and has only a finite number of physical gauge invariant operators.
In other words, what we are really exploring here is the
non-gauge-fixed level of the theory, which describes the gravitational
field as seen by a physical reference system \cite{obs}.

\item We do not measure lengths directly as numbers: numbers are given
by ratios between physical lengths.  For instance, by the number of
times a rod fits into an interval.  One may thus wonder whether the
sign or the imaginary character of the interval has any importance by
itself.  The answer is of course not.  The imaginary unit simply keeps
track of the distinction between the two kind of intervals, which are
fundamentally distinguished from each other by their relations, namely
by the different way in which they fit into a Minkowski (or a locally
Minkowskian) space.  It is interesting to notice that these relations
between intervals are in fact reproduced by the $su(1,1)$
representation theory.  Spacelike and timelike intervals sum among
themselves differently, and this is reflected in the way direct
products of representations can be decomposed.  This works if we
identify timelike intervals with discrete representation (plus or
minus, according to future and past) and spacelike ones with
continuous representations.  This is illustrated in the previous
section and the Appendix.  For
instance, the sum of two future timelike vectors can only be a future
timelike vector.  Accordingly, the direct sum of two representations
in the ${\cal D}^+_n$ series contains only representations of the
${\cal D}^+_n$ series.  This fact reinforces the idea that the
discrete representations are naturally timelike and the continuous
ones are ``naturally" spacelike.

\item
Using the correspondence between Chern-Simons theory and 3
dimensional gravity, we expect the introduction of a cosmological
constant $\Lambda>0$
to deform the group structure to a quantum group structure
$U_{q}(SU(1,1))$, with the deformation parameter $q$ being related to the
cosmological length $L=1/\sqrt{\Lambda}$.
The representations of interest are still discrete or continuous. The novelty
is that the continuous representations for $q=e^{-h}$
admit an infrared cutoff given by $\f{\pi}{2h}$ \cite{suq},
so that  no spacelike length can be bigger than $L=\f{\pi}{2h}L_P$, which is
identified to the cosmological scale or, in other words,
the distance to the cosmological horizon.
This is consistent with the physical intuition
that no information is accessible behind the horizon.
This way, we expect a relation of the type
$q\sim e^{-\sqrt{\Lambda}}$, or more precisely
$q=e^{-\f{\pi L_P}{2L}}$.
Similar considerations have recently been developed in a 4-dimensional
spin foam approach \cite{nouiroche}.

\item
We have introduced two length operators.
The first one corresponds to the usual way of quantizing geometrical
operators in loop quantum gravity. The spectrum of this operator 
has both a minimal timelike distance 
{\it and} a minimal spacelike distance. This may be surprising, given that 
the spacelike spectrum is continuous.
Also, representations corresponding to the null directions 
appear in the spectrum of this operator.\footnotemark
\footnotetext{
It might be interesting to notice that using Hod's \cite{Hod} correspondence
principle between quasi normal
modes and quantization of black hole area fluctuations, it has been
recently argued \cite{Olaf} that the minimum allowed spin relevant for
the black hole horizon might be $1$. This 
might perhaps be related with the fact that the spin 1 representation 
correspond to null directions.} 
Finally, the timelike spectrum is not equally spaced.\footnotemark
\footnotetext{On the connection between the fact that the spectrum of the area is not
equally spaced and the Hawking thermal radiation see \cite{bh}.}
On the other hand, the second operator issued from the symmetric
quantization map, agrees with the spin foam computations and the
radius of the coadjoint orbits. In this case 
the timelike spectrum is discrete (there is a minimal length) and
equally spaced (moreover, when using $SU(1,1)$ instead of $SO(2,1)$,
we allow spin $n\in \N+1/2$ and the difference between two consecutive
length values becomes exactly the minimum allowed length),
and the spacelike spectrum is continuous and has no
initial gap. 

\item As we mentioned at the end of the introduction, the result in
this letter shows that in 3d there is consistency between spin foam
\cite{davids} and loop results.  The situation is still unclear in 4d,
where there is an apparent sign discrepancy between spin foam
\cite{BC,alex} and loop \cite{discrete1,discrete2} results.  
In 4d, so far the focus has mostly been on the absolute
value, and not on the sign, of quantum geometrical quantities; 
a detailed investigation of the signature of the area in the
quantum regime, and a careful comparison of the spin foam and canonical
results, would be of interest.  The analysis of the  covariant canonical
structure of general relativity recently
completed in \cite{richard} might be a useful step in this direction.

\end{itemize}

\appendix
\section{Representations of $SU(1,1)$}
\label{su11}

Here we  extend  the results on the geometrical operators to the
group $SU(1,1)$, the double cover of $SO(2,1)$. Just as when
extending $SO(3)$ to $SU(2)$, this extension  doubles the number of
representations and introduces a parity.

In the principal series of $SU(1,1)$,
there are two series of continuous representations ${\cal
C}^\epsilon_s$ where $\epsilon=0,1/2$ is the parity and $s$ a positive
real number.  The Casimir is $q=s^2 +1/4 >0$ and the set of weights
$m$ is formed by the integers or the half-integers depending on the parity of
the representation.  There are two series of discrete
representations ${\cal D}^\pm_n$ labelled by a half-integer $n$
larger than 1.  The Casimir is $q=n(1-n)<0$ and the set of weights $m$
is $n+\N$ for the positive series, and
$-(n+\N)$ for the negative one.
The Plancherel formula for a function $f\in L^2(SU(1,1))$ reads:
\beqs
f(g)&=&\int_{s>0} ds\ \frac{\coth(\pi s)}{4\pi s}\
\textrm{Tr}(f^0_sR_{{\cal C}^0_s}(g)) +\int_{s>0} ds\ \frac{\tanh(\pi
s)}{4\pi s}\ \textrm{Tr}(f^0_sR_{{\cal C}^{1/2}_s}(g))  \nonumber\\
&&+\sum_{n\ge 1}\ (2n-1)\ \textrm{Tr}(f^+_nR_{{\cal D}^+_n}(g))
+\sum_{n\ge 1}\ (2n-1)\ \textrm{Tr}(f^-_nR_{{\cal D}^-_n}(g)).
\eeqs
The Casimir (shifted by one fourth) still give the (square of) length
associated to the edge labelled by the representation.
The representations $n=\f{1}{2}^\pm$ are unitary but do not enter the
Plancherel decomposition. Physically, their corresponding length is 0
and they correspond to null edges.

As for the area operator and the admissible nodes, there is not
much that changes. The only difference is that one must take care of the
parity. The tensor product decomposition reads:
\begin{equation}
    {\cal D}^\pm_{n_1} \otimes {\cal C}_{s_2}^{\epsilon_2} =
    \bigoplus_{n\ge n_{min}} {\cal D}^\pm_{n} \oplus
    \int ds \ {\cal C}_s^\epsilon,
\end{equation}
\begin{equation}
    {\cal C}_{s_1}^{\epsilon_1} \otimes {\cal C}_{s_2}^{\epsilon_2}=
    \bigoplus_{n\ge n_{min}} {\cal D}^+_{n} \oplus
    \bigoplus_{n\ge n_{min}} {\cal D}^-_{n} \oplus
    2\int ds\  {\cal C}_s^\epsilon
\end{equation}
\begin{equation}
    {\cal D}^\pm_{n_1}\otimes{\cal D}^\pm_{n_2}=
    \bigoplus_{n\ge n_1+n_2}{\cal D}^\pm_n,
\end{equation}
\begin{equation}
    {\cal D}^+_{n_1}\otimes{\cal D}^-_{n_2}=
    \bigoplus_{n=n_{min}}^{n_1-n_2}{\cal D}^+_{n} \oplus
    \bigoplus_{n=n_{min}}^{n_2-n_1}{\cal D}^-_{n} \oplus
    \int ds \ {\cal C}_s^\epsilon.
\end{equation}
where $n_{min}=1$ and $\epsilon=0$ if $n_1+n_2$ is an integer,
$n_{min}=3/2$ and $\epsilon=1/2$ otherwise.


\begin{thebibliography}{99}

\bibitem{discrete1}
C Rovelli, L Smolin: {\em Discreteness of Area and Volume in Quantum
Gravity,\/} Nucl Phys B 442 (1995) 593.  Erratum: Nucl Phys B 456 (1995)
734.

\bibitem{discrete2}
A Ashtekar, J Lewandowski, {\em Quantum Theory of Gravity I: Area
Operators,\/} Class Quantum Grav 14 (1997) A55.

\bibitem{ponzano}
C Rovelli, {\em Basis of the Ponzano-Regge-Turaev-Viro-Ooguri quantum
gravity model is the loop representation basis,\/} Phys Rev D48
(1993) 2702.

\bibitem{sergei}
S Alexandrov,
{\em $SO(4,C)$-covariant Ashtekar-Barbero gravity and the Immirzi parameter},
gr-qc/0005085,
Class.Quant.Grav. 17 (2000) 4255-4268

\bibitem{sergei&dima}
S Alexandrov, D Vassilevich,
{\em Area spectrum in Lorentz covariant loop gravity},,
gr-qc/0103105,
Phys.Rev. D64 (2001) 044023

\bibitem{richard}
S Alexandrov, ER Livine,
{\it SU(2) Loop Quantum Gravity seen from Covariant Theory},
gr-qc/0209105

\bibitem{loop}
C Rovelli, L Smolin, Phys Rev Lett 61 (1988) 1155; Nucl Phys B331
(1990) 80.  A Ashtekar, J Lewandowski, D Marolf, J Mourao, T Thiemann,
J Math Phys 36 (1995) 6456.  For overview and references on loop
quantum gravity: C Rovelli: {\em Loop Quantum Gravity,\/} Living Reviews
in Relativity, http://www.livingreviews.org/Articles/Volume1/%
1998-1rovelli; gr-qc/9709008.  For an introduction, in particular to
techniques used here: M Gaul, C Rovelli, {\em Loop Quantum Gravity and
the Meaning of Diffeomorphism Invariance,\/} in ``Towards Quantum
Gravity", J Kowalski-Glikman ed (Springer Lecture Notes in Physics,
Berlin, 2000), gr-qc/9910079.

\bibitem{orbit}
E Witten, {\it Coadjoint Orbits of the Virasoro Group,\/}
Commun Math Phys 114 (1988) 1-53.

\bibitem{BC}
JW Barrett, L Crane, {\em A Lorentzian Signature Model for Quantum
General Relativity,\/} gr-qc/9904025, Class Quant Grav 17 (2000)
3101-3118.


\bibitem{alex}
A Perez, C Rovelli: {\em Spin foam model for Lorentzian General
Relativity,\/} gr-qc/0009021; {\em 3+1 spinfoam model of quantum
gravity with spacelike and timelike components,\/} gr-qc/0011037.

\bibitem{hooft}
G 't Hooft,
{\em Canonical Quantization of Gravitating Point Particles in 2+1
dimensions},
gr-qc/9305008,
Class.Quant.Grav. 10 (1993) 1653-1664

\bibitem{lpr}
L Freidel,
{\it A Ponzano-Regge model of Lorentzian 3-Dimensional gravity},
Nucl.Phys.Proc.Suppl. 88 (2000) 237-240,
gr-qc/0102098

\bibitem{davids}
S Davids,
{\it A State Sum Model for (2+1) Lorentzian Quantum Gravity},
gr-qc/0110114

\bibitem{larea}
S Alexandrov,
{\em On choice of connection in loop quantum gravity},
gr-qc/0107071,
Phys.Rev. D65 (2002) 024011

\bibitem{jacek}
J Wisniewski,
{\em 2+1 General Relativity: Classical and Quantum},
PhD Thesis (PennState University, 2002)

\bibitem{thiemann3d}
T Thiemann,
{\it QSD IV : 2+1 Euclidean Quantum Gravity as a model to test 3+1 Lorentzian Quantum Gravity},
Class.Quant.Grav. 15 (1998) 1249-1280,
gr-qc/9705018

\bibitem{spinnet}
L Freidel, ER Livine,
{\it Spin networks for Non-Compact groups},
hep-th/0205268,
to appear in J.Math.Phys.

\bibitem{primer}
C Rovelli, P Upadhya, {\it Loop quantum gravity and quanta of space: a
primer}, gr-qc/9806079

\bibitem{jerzy}
J Lewandowski, ET Newman, C Rovelli, {\it Variation of the parallel
propagator and holonomy operator and the Gauss law constraint},
J Math Phys 34 (1993) 4646.

\bibitem{alekseev}
A Alekseev, AP Polychronakos, M Smedb{\"a}ck,
{\it On area and entropy of a black hole},
hep-th/0004036

\bibitem{sf}
L Freidel, K Krasnov
{\it Spin Foam Models and the Classical Action Principle},
Adv.Theor.Math.Phys. 2 (1999) 1183-1247,
hep-th/9807092

\bibitem{louapref}
L Freidel, D Louapre,
{\it  Asymptotics of 6j and 10j symbols},
hep-th/0209134

\bibitem{matter}
C Rovelli, {\em A generally covariant quantum field theory and a
prediction on quantum measurements of geometry,\/} Nucl Phys B405,
797 (1993).

\bibitem{matter2}
L Smolin, {\em Finite, diffeomorphism invariant observables
in quantum gravity,\/} Phys Rev  D49 (1994) 4028-4040.

\bibitem{obs}
C Rovelli, {\em What is observable in classical and quantum
gravity?,\/} Class Quantum Grav 8 (1991) 297.

\bibitem{suq}
Y Masuda, K Mimachi, Y Nakagami, M Noumi, Y Saburi, K Ueno,
Lett.Math.Phys 19 (1990) 187.

\bibitem{nouiroche} K Noui, P Roche
{\it Cosmological Deformation of Lorentzian Spin Foam Models},
 gr-qc/0211109.

\bibitem{Hod} S Hod,
{\it Gravitation, the Quantum, and Bohr's Correspondence
    Principle},
Gen. Rel. Grav. 31 (1999) 1639, gr-qc/0002002.

\bibitem{Olaf} O Dreyer,
{\it Quasinormal Modes, the Area Spectrum, and Black Hole Entropy},
gr-qc/0211076.

\bibitem{bh}
M Barreira, M Carfora, C Rovelli,
{\em 
Physics with nonperturbative quantum gravity: radiation from a quantum black hole},
gr-qc/9603064,
Gen.Rel.Grav. 28 (1996) 1293-1299



\end{thebibliography}
\end{document}